\documentclass[twocolumn,showpacs,preprintnumbers,amsmath,amssymb,superscriptaddress]{revtex4}

\usepackage{graphicx}
\pdfoutput=1
\begin{document}
\bibliographystyle{prsty}
\title{Origin of the anomalous mass renormalization in metallic quantum well states of correlated oxide SrVO$_3$
}
\pdfoutput=1

\author{Masaki Kobayashi}
\altaffiliation{Author to whom correspondence should be addressed; masakik@post.kek.jp}
\affiliation{Photon Factory, Institute of Materials Structure Science, 
High Energy Accelerator Research Organization (KEK), 
1-1 Oho, Tsukuba 305-0801, Japan}
\author{Kohei Yoshimatsu}
\altaffiliation{Present address: Department of Applied Chemistry, Tokyo Institute of Technology, 
Tokyo 152-8522, Japan}
\affiliation{Photon Factory, Institute of Materials Structure Science, 
High Energy Accelerator Research Organization (KEK), 
1-1 Oho, Tsukuba 305-0801, Japan}
\affiliation{Department of Physics, University of Tokyo, 
7-3-1 Hongo, Bunkyo-ku, Tokyo 113-0033, Japan}
\author{Enju Sakai}
\affiliation{Photon Factory, Institute of Materials Structure Science, 
High Energy Accelerator Research Organization (KEK), 
1-1 Oho, Tsukuba 305-0801, Japan}
\author{Miho Kitamura}
\affiliation{Photon Factory, Institute of Materials Structure Science, 
High Energy Accelerator Research Organization (KEK), 
1-1 Oho, Tsukuba 305-0801, Japan}
\author{Koji Horiba}
\affiliation{Photon Factory, Institute of Materials Structure Science, 
High Energy Accelerator Research Organization (KEK), 
1-1 Oho, Tsukuba 305-0801, Japan}
\author{Atsushi Fujimori}
\affiliation{Department of Physics, University of Tokyo, 
7-3-1 Hongo, Bunkyo-ku, Tokyo 113-0033, Japan}
\author{Hiroshi Kumigashira}
\affiliation{Photon Factory, Institute of Materials Structure Science, 
High Energy Accelerator Research Organization (KEK), 
1-1 Oho, Tsukuba 305-0801, Japan}

\date{\today}

\begin{abstract}
{\it In situ} angle-resolved photoemission spectroscopy (ARPES) has been performed on SrVO$_3$ ultrathin films, which show metallic quantum well (QW) states, to unveil the origin of the anomalous mass enhancement in the QW subbands. 
The line-shape analysis of the ARPES spectra reveals that the strength of the electron correlation increases as the subband bottom energy approaches the Fermi level. 
These results indicate that the anomalous subband-dependent mass enhancement mainly arises from the quasi-one-dimensional character of confined V $3d$ states as a result of their orbital-selective quantization.
\end{abstract}


\maketitle

Quantum confinement of strongly correlated electrons in oxide heterostructures has attracted considerable interest not only for its potential for technological applications in future oxide electronics, but also for the opportunity it presents to better understand the fundamental low-dimensional physics of strongly correlated electron systems \cite{NatMater_12_Hwang}. 
The reduction of the dimensionality changes the complex interaction among the spin, charge, and orbital degrees of freedom of the correlated electrons, resulting in the emergence of unusual quantum phenomena \cite{Science_10_Takagi, Science_10_Mannhart}. 
Recently, metallic quantum well (QW) states have been clearly observed by using angle-resolved photoemission spectroscopy (ARPES) for strongly correlated electrons in SrVO$_3$ (SVO) ultrathin films grown on SrTiO$_3$ (STO) substrates, where the high-density strongly correlated electrons of the order of $\sim 10^{22}$ cm$^{-3}$ are confined in a narrow space on the scale of a few nanometers \cite{Science_11_Yoshimatsu}. 
This situation is distinct from the two-dimensional electron gas (2DEG) states in oxide semiconductors \cite{Nature_04_Ohtomo, NatMater_10_Tsukazaki, Nature_11_Santander-Syro, NatMater_11_Meevasana, PRB_12_Santander-Syro, NatComm_14_King, PRL_14_Plumb}. 
Thus, such a use of quantum confinement to form two-dimensional electron liquid (2DEL) states will provide a foundation for studying the behavior of strongly correlated electrons under reduced dimensions and for controlling the extraordinary physical properties of strongly correlated oxides.

The observed metallic QW states in SVO ultrathin films exhibit two distinctive features \cite{Science_11_Yoshimatsu}. 
The first is the {\it orbital-selective quantization} originating from the anisotropic orbital character of the V $3d$ $t_{2g}$ ($d_{xy}$, $d_{yz}$, and $d_{zx}$) band states. 
In bulk crystals, each band essentially has a two-dimensional (2D) character in the $xy$, $yz$, and $zx$ planes, respectively. 
In thin films, the 2D nature of each band causes the further reduction of dimensionality to one dimension that is determined by the quantization direction: When the SVO film has become sufficiently thin in the $z$ direction to realize quantum confinement, the bands derived from the $d_{yz}/d_{zx}$ orbitals are subject to quantization (changing from 2D to 1D (one-dimensional)), while the $d_{xy}$ band states remain unchanged (maintaining their 2D nature). 
The other distinctive feature is the {\it anomalous mass enhancement}, which depends on the subbands: The subband dispersion becomes considerably narrower as the subband bottom energy ($E_n$) approaches the Fermi level ($E_\mathrm{F}$). 
Such anomalous subband-dependent mass enhancement has not been observed in conventional metallic QW structures based on metals having nearly free-electron-like $sp$ bands \cite{PRB_90_Mueller, PRB_02_Matsuda}, suggesting the importance of underlying strongly correlated electronic states in the SVO QW.

The behavior observed in SVO QWs cannot be understood within a simple Fermi liquid picture where electron-electron scattering depends only on the electron energy measured from $E_\mathrm{F}$: In that picture, electron-electron correlation in the subband should be reduced as the band filling of the subband is reduced; that is, the subband bottom energy approaches $E_\mathrm{F}$ owing to the reduction in band filling. 
Okamoto has studied the subband structures of correlated QWs of SVO by using a layered dynamical-mean-field theory (DMFT) calculation, and noted the importance of both the {\it band effect}, originating from the long-range hopping, and the {\it electron-electron correlation}, due to short-range Coulomb repulsion \cite{PRB_11_Okamoto}. 
Because the band effect gives rise to the apparent enhancement of the effective mass in subbands of higher quantum numbers even without electron correlation, the effect of electron-electron correlation on the subband narrowing is still not understood \cite{PRB_11_Okamoto, PRB_13_Zhong}. 
Therefore, the experimental evaluation of the electron correlation strength in the subband from a different perspective is indispensable for revealing the origin of the anomalous mass enhancement.

In this Letter, we report the results of {\it in situ} ARPES measurements of metallic QW structures formed from SVO and a detailed line-shape analysis of the ARPES spectra, with particular focus on the widths of the momentum distribution curves (MDCs). 
The spectral analysis using MDCs enables us to distinguish between the effect of electron correlation and the band effect in the observed subband narrowing. 
The line-shape analysis demonstrates the significant enhancement of the self-energy as the subband bottom energy approaches $E_\mathrm{F}$, indicating the importance of electron correlation in the QW states. 
If the electron correlation is dominated by intra-subband electron-electron scattering, the dependence of the strength increase of the electron correlation on the subband can be explained by the quasi-1D nature of the orbital-selective quantized states in SVO.

Digitally controlled SVO ultrathin films were grown on the atomically flat (001) surface of TiO$_2$-terminated Nb-doped STO substrates in a laser molecular-beam epitaxy chamber connected to an ARPES system at BL-28 of the Photon Factory (PF) \cite{RSI_03_Horiba}. 
During the growth of an SVO film, the thickness was precisely controlled on the atomic scale by monitoring the intensity oscillation of reflection high-energy electron diffraction (RHEED). The details of the growth conditions are described elsewhere \cite{PRL_10_Yoshimatsu, Science_11_Yoshimatsu}. 
The characterization results were essentially the same as those reported in the previous studies \cite{PRL_10_Yoshimatsu, Science_11_Yoshimatsu}. 
The films prepared for the measurements were transferred under an ultrahigh vacuum of 10$^{-10}$ Torr to the photoemission chamber to avoid the degradation of the SVO surface upon exposure to air. 
The ARPES experiments were conducted {\it in situ} under an ultrahigh vacuum of 10$^{-11}$ Torr at a temperature ($T$) of 20 K using horizontal linear polarization of the incident light. 
The energy and angular resolutions were respectively set to about 30 meV and 0.3$^{\circ}$. 
The Fermi level of the samples was calibrated by measuring a gold foil that was electrically connected to the samples. The details of the ARPES measurement setups are described elsewhere \cite{Science_11_Yoshimatsu}.

\begin{figure}[t!]
\begin{center}
\includegraphics[width=8.8cm]{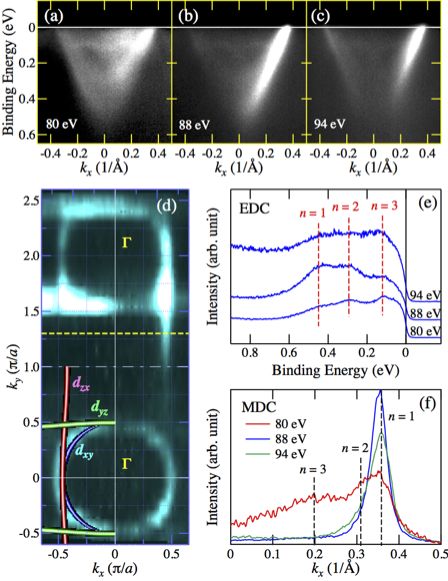}
\caption{(color online). 
Photon-energy dependence of ARPES spectra for the 7-ML SrVO$_3$ ultrathin film. 
(a)-(c) ARPES images taken at $h\nu = 80$ eV, 88 eV, and 94 eV, respectively. 
The ARPES spectra were measured along the $k_x$ slice at $k_y = 1.3 \pi/a$ (indicated by the dashed line in (d)), where the spectra include only the quantized $d_{zx}$ subbands. 
(d) Fermi surface mapping measured at $h\nu = 88$ eV, where the $n = 1$ states are observed to be dominant. The results of tight-binding calculations for the $n = 1$ states are superimposed as solid curves. 
The dashed line represents the measurement line for (a)-(c), and the dot-dashed line is the Brillouin zone boundary. 
(e) Energy distribution curves (EDCs) at the point (0, $1.3 \pi/a$). 
The vertical dashed lines denote the subband energy minima $E_n$ of the QW states. 
(f) Momentum distribution curves (MDCs) at $E_\mathrm{F}$. 
The vertical dashed lines denote the $k_\mathrm{F}$ values of the QW states. 
The values of $k_\mathrm{F}$ for the $n=1$, 2, and 3 states are 0.20, 0.31, and 0.35 \AA$^{-1}$, respectively. 
}
\label{hvDep}
\end{center}
\end{figure}

Figures \ref{hvDep}(a)-\ref{hvDep}(c) show ARPES images of the 7-monolayer (ML) SVO ultrathin films taken at various photon energies ($h\nu$) along the cut indicated by the dashed line in the Fermi surface (FS) mapping of Fig.~\ref{hvDep}(d). 
In comparison with the prediction given by the tight-binding (TB) calculations, a circular FS centered at the $\Gamma$ point is derived from the $d_{xy}$ band, while the two nearly parallel linear FSs are from the $d_{yz}$ and $d_{zx}$ bands \cite{M.E.}. 
Since the band dispersions along the cut are derived only from the quantized $d_{zx}$ band states, the ARPES spectra consist of the three $d_{zx}$-derived subbands with quantum numbers $n = 1$, 2, and 3 \cite{Science_11_Yoshimatsu}. 
The ARPES images demonstrate that the spectral intensity of each quantized band significantly depends on $h\nu$ \cite{PRL_80_Louie}. 
Although the intensities strongly change with $h\nu$, the subband dispersions themselves remain unchanged owing to the low-dimensional character of the quantized states. 
The 2D nature of the subbands is also confirmed by the energy distribution curves (EDCs) taken at the (0, $1.3 \pi/a$) point, as shown in Fig.~\ref{hvDep}(e). 
For the MDCs in Fig.~\ref{hvDep}(f), a much stronger $h\nu$-dependent intensity modulation is observed around the Fermi momentum ($k_\mathrm{F}$). 
As expected from the ARPES images in Figs.~\ref{hvDep}(b) and \ref{hvDep}(c), the intensity of the $n = 1$ subband is too strong to allow it to be compared with the other subbands ($n = 2$ and 3) in the ARPES images measured with $h\nu = 88$ eV and 94 eV, i.e., the peak structures corresponding to the $n = 2$ and 3 band states are buried in the tail of the prominent $n = 1$ states. 
In contrast, the intensity of the $n = 1$ subband is strongly suppressed and becomes comparable to those of the higher $n$ subbands in the MDC taken at $h\nu = 80$ eV. 
It should be noted that the $h\nu$ dependence of the spectral intensity shows almost the same behavior irrespective of the thickness of the SVO ultrathin films. 
Therefore, the ARPES data taken with $h\nu = 80$ eV are used in the detailed line-shape analysis discussed below.

\begin{figure}[t!]
\begin{center}
\includegraphics[width=8.8cm]{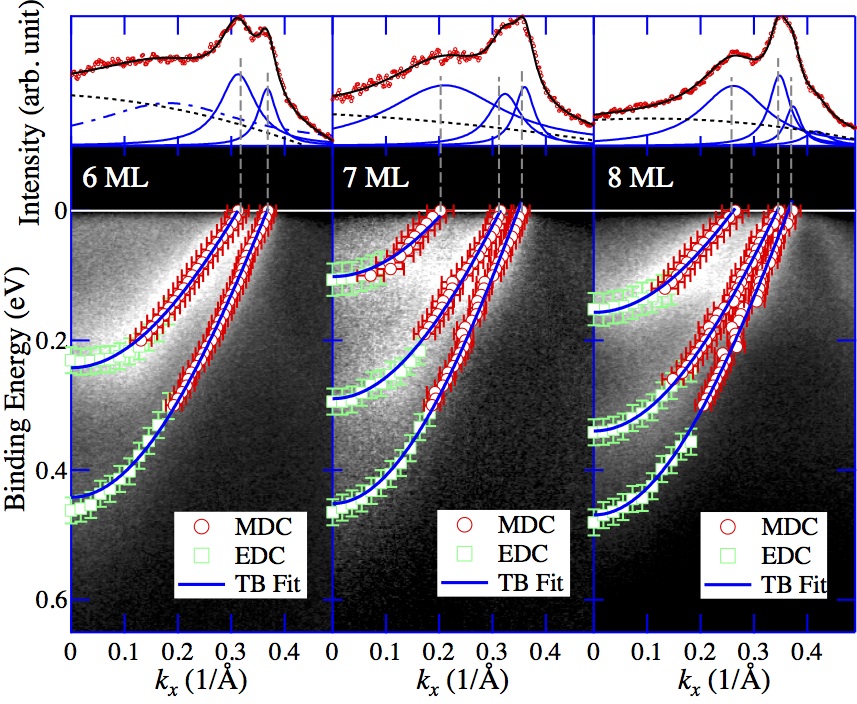}
\caption{(color online). 
Line-shape analysis for SrVO$_3$ QW states. 
The intensity plots of ARPES spectra for the 6-ML (a), 7-ML (b), and 8-ML (c) SrVO$_3$ films are obtained at the point (0, $1.3 \pi/a$). 
The top panels show the MDCs at $E_\mathrm{F}$ within an energy window of $\pm 10$ meV. 
The MDCs have been fitted to a linear combination of Lorentzians with a smooth background. 
The dispersions of the subbands have been determined from the MDC peak positions (open circles) in the vicinity of $E_\mathrm{F}$ and the EDC peak positions (open squares) near the minima of the subbands, respectively. 
The solid curves represent the curve fits (based on the tight-binding model Eq.~(\ref{e(k)TB})) to the dispersion of the QW states. 
The dashed lines are the backgrounds. 
The broad hump-like structure around the $\Gamma$ point (dashed-dotted line) in the 6-ML film may be due to the contribution from the $n = 3$ state at the unoccupied states in the vicinity of $E_\mathrm{F}$. 
}
\label{E-k}
\end{center}
\end{figure}

Figure~\ref{E-k} the ARPES intensity plots for 6-8 ML SVO ultrathin films taken along the same $k_x$ slice at $k_y = 1.3 \pi/a$ indicated by a dashed line in Fig.~\ref{hvDep}(d), together with the peak positions determined from the MDCs for the near-$E_\mathrm{F}$ region and the EDCs for the energy regions near the minima of the subbands. The MDCs are reproduced by the combination of Lorentzians corresponding to the respective subbands with a smooth background \cite{Science_99_Valla}. 
In the top panels, the MDCs at $E_\mathrm{F}$ are shown for the corresponding SVO ultrathin films. 
It is evident that the width of the MDC peak ($\Delta k$) becomes wider with increasing quantum number $n$ for each film thickness and as the bottom energy approaches $E_\mathrm{F}$, suggesting that the electron correlation makes a siginificant contribution to the subband-dependent mass enhancement. 
In order to estimate the ``renormalization factor'' $Z_n^\mathrm{TB}$ in accordance with the previous study \cite{Science_11_Yoshimatsu}, a fit to the dispersion of each subband, $\varepsilon_n (\mathbf{k})$, is performed, using the equation:
\begin{equation}
\varepsilon_n (k_\parallel) = Z_n^{\mathrm{TB}} \, \varepsilon^{\mathrm{TB}}(k_\parallel) + \varepsilon' ,
\label{e(k)TB}
\end{equation}
where $k_\parallel$ is the in-plane momentum, $\varepsilon^\mathrm{TB} (k_\parallel)$ is the bulk band dispersion calculated using the tight-binding parameters of bulk SVO \cite{NJP_05_Pavarini}, and $\varepsilon'$ is a parameter that can be adjusted so as to match the minima of the subbands with experimental results. 
As can be seen in Fig.~\ref{E-k}, the observed subband dispersions are well reproduced by Eq.~(\ref{e(k)TB}), as is the FS topology shown in Fig.~\ref{hvDep}(d).

According to the theoretical analysis based on the layered DMFT \cite{PRB_11_Okamoto}, the obtained $Z_n^\mathrm{TB}$ is expressed as the product of factors representing the band effect $Z_n^\mathrm{band}$ and the electron correlation effect $Z_n^{e-e}$. 
Thus, a crucial issue is how the net effect of electron correlation on the subband narrowing is extracted from the ARPES data, since both the band effect and the electron correlation effect give rise to the narrowing of the subband dispersion. 
To shed some light on the influence of the electron correlation on the mass enhancement, we have employed the line-shape analysis of MDCs as a function of binding energy $\omega$. 
The width $\Delta k$ of the peak in an MDC reflects the imaginary part of the {\it self-energy}, $\mathrm{Im} \Sigma (\mathbf{k}, \omega)$, which is related to the inverse lifetime $1/\tau$, as follows \cite{Science_99_Valla}:
\begin{equation}
\left| 2 \mathrm{Im} \Sigma (\mathbf{k}, \omega) \right| \approx \hbar v_k \Delta k = \frac{\hbar}{\tau}.
\label{ImS}
\end{equation}
Here, $\hbar$ is the Planck constant and $v_k$ is the velocity along the band dispersion. 
Using this equation, one can deduce the imaginary part of the self-energy, $\mathrm{Im}\Sigma (\mathbf{k}, \omega)$, from the measured $\Delta k$ values. 
Thus, the spectral analysis using the MDC enables us to extract the effect of electron correlation irrespective of the apparent subband narrowing due to the band effect.

\begin{figure}[t!]
\begin{center}
\includegraphics[width=8.8cm]{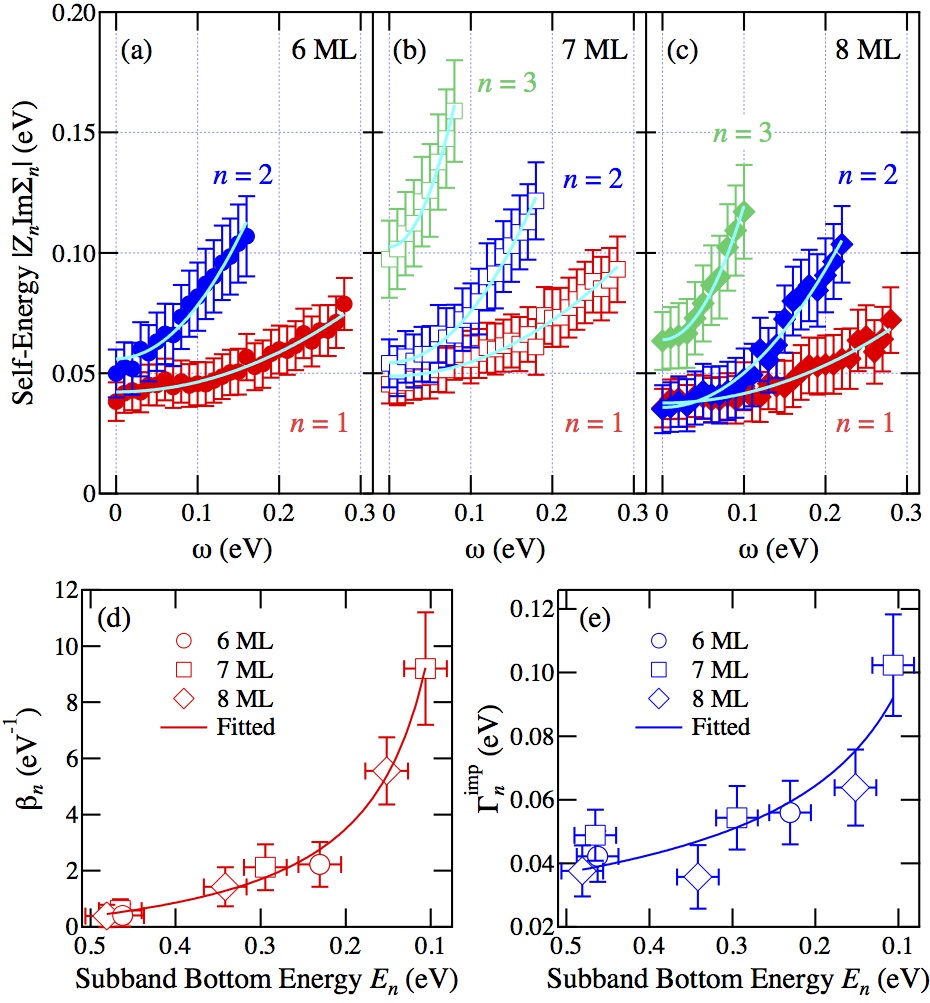}
\caption{(color online). 
Self-energy $\Sigma$ for SrVO$_3$ QW states. 
(a)-(c) Imaginary part of the self-energy, $Z_n \mathrm{Im}\Sigma_n (\omega)$, of each subband plotted as a function of $\omega$ for 6-ML, 7-ML, and 8-ML SVO films. 
The solid curves represent the fitted curves based on Eq.~(\ref{ImS4FL}). 
(d), (e) Respectively the $E_n$ dependences of the coefficients $\beta_n$ and $\Gamma_n^\mathrm{imp}$ estimated from the curve fits. 
The solid curves are the results of curve fitting assuming the quasi-1D density of states: 
$\beta_n \propto 1/E_n$ and $\Gamma_n^\mathrm{imp} \propto 1/\sqrt{E_n}$. For details, see the text. 
}
\label{ImS_beta}
\end{center}
\end{figure}

In a previous ARPES study on thick SVO films (with thicknesses greater than 100 ML) \cite{PRL_12_Aizaki}, the self-energy $\Sigma (\mathbf{k}, \omega)$ showed a typical Fermi liquid behavior, reflecting the correlated Fermi-liquid ground states of bulk SVO \cite{PRB_98_Inoue}. 
This fact leads to the reasonable assumption that the QW states also exhibit a correlated Fermi-liquid behavior. 
Another point to note is that $\mathrm{Im}\Sigma (\mathbf{k}, \omega)$ in SVO films is nearly identical between cuts across different Fermi momenta, indicating that the self-energy $\Sigma$ is particularly independent of $\mathbf{k}$ \cite{PRL_12_Aizaki}. 
If we neglect the $\mathbf{k}$-dependence of the self-energy for the quantized subbands as well, the imaginary part of the self-energy in the vicinity of $E_\mathrm{F}$ for each of the subbands is expressed as \cite{RMP_98_Imada}:
\begin{equation}
\left| 2 Z_n \mathrm{Im} \Sigma_n (\omega) \right| \approx \hbar v_\mathrm{F}^n \Delta k_n,
\label{ZnImSn}
\end{equation}
where $v_\mathrm{F}^n = \frac{1}{\hbar} \left( \frac{d\varepsilon_n (\mathbf{k})}{d k_\parallel} \right)$ is the Fermi velocity determined from experimental data. 
The energy ranges for the application of Eq.~(\ref{ZnImSn}) are defined so as to maintain the linearity of each subband dispersion from $E_\mathrm{F}$. 
The estimated $Z_n \mathrm{Im}\Sigma_n (\omega)$ values for each of the QW subbands are shown in Figs.~\ref{ImS_beta}(a)-\ref{ImS_beta}(c). 
Each $\mathrm{Im}\Sigma_n (\omega)$ curve monotonically increases with increasing $\omega$. 
Although electron-phonon scattering may weakly affect $\mathrm{Im}\Sigma_n (\omega)$ in the region of $\sim 60$ meV below $E_\mathrm{F}$ as observed in the case of thick SVO films \cite{PRL_12_Aizaki}, no clear signature indicating this effect is obtained in any of the $\mathrm{Im}\Sigma_n (\omega)$ curves, suggesting that the effect of the electron-phonon scattering on each $\mathrm{Im}\Sigma_n (\omega)$ can be ignored in the present experiment. 
Furthermore, the rate of increase of $\mathrm{Im}\Sigma_n (\omega)$ as a function of $\omega$ tends to be larger with increasing $n$ and/or decreasing film thickness, namely as the minimum of the subband approaches $E_\mathrm{F}$, clearly demonstrating the subband-dependent renormalization due to electron correlation.

In order to evaluate the electron correlation strength for each of the subbands quantitatively, the $\mathrm{Im}\Sigma_n (\omega)$ curves are fitted to the following equation, which assumes the Fermi-liquid ground states:
\begin{equation}
\left| Z_n \mathrm{Im} \Sigma_n (\omega) \right| = \Gamma_n^\mathrm{imp} + \beta_n \left( \omega^2 + \left( \pi k_\mathrm{B} T \right)^2 \right),
\label{ImS4FL}
\end{equation}
where $\beta_n$ denotes a coefficient reflecting the strength of electron-electron interaction, $k_\mathrm{B}$ is the Boltzmann constant, and $\Gamma_n^\mathrm{imp}$ is the inverse lifetime of quasi-particles associated with the impurity scattering. 
The obtained values of $\beta_n$ and $\Gamma_n^\mathrm{imp}$ are plotted against the bottom energy of subband, $E_n$, in Figs.~\ref{ImS_beta}(d) and \ref{ImS_beta}(e), respectively. 
As $E_n$ approaches $E_\mathrm{F}$, these values are considerably enhanced, and the enhancement seems to be dependent only on $E_n$.

Now we examine the relationship between the renormalization factor $Z_n^{e-e}$ and the coefficient $\beta_n$, in order to check the validity of the present MDC analysis for mass renormalization. 
In the Fermi-liquid states, the well-known Kadowaki-Woods relation states that the ratio between $A$ and $\gamma^2$ is a universal constant \cite{SSC_86_Kadowaki}, where $A$ is the coefficient of the $T^2$ term in the resistivity and $\gamma$ is for the $T$-linear term in the electric specific heat coefficient. 
The renormalization factor $Z^{e-e}$ is related to the real part of the self-energy, $\mathrm{Re}\Sigma(\omega)$; the first-order approximation for $\mathrm{Re}\Sigma(\omega)$ in the vicinity of $E_\mathrm{F}$ is $\mathrm{Re}\Sigma(\omega) \approx - \alpha \omega$, and consequently the bandwidth near $E_\mathrm{F}$ becomes narrowed by a factor of $Z^{e-e} = \left[ 1+ \alpha \right] ^{-1}$. 
Then, the quasiparticle density of states at $E_\mathrm{F}$ is enhanced by $1/Z^{e-e}$, and the quantity $1/Z^{e-e}$ is proportional to the electronic specific-heat coefficient $\gamma$. 
Meanwhile, $A$ is proportional to $\beta$, as is clear from Eq.~(\ref{ImS4FL}). 
Considering the Kadowaki-Woods relation, the following relationship is derived:
\begin{equation}
\beta_n \times \left( Z_n^{e-e} \right)^2 \propto A/\gamma^2 = \mathrm{const.},
\label{KWeq}
\end{equation}
indicating that the value of $\beta_n$ is related to the mass renormalization of the subbands ($1/Z_n^{e-e}$). 
Therefore, the determination of $\beta_n$ corresponds to the evaluation of the contribution of the electron correlation in each subband and is free from the influence of the band effect. 
As can be seen from Fig.~\ref{ImS_beta}(d), the estimated $\beta_n$ monotonically increases with decreasing $E_n$, providing the experimental evidence that the electron correlation significantly contributes to the anomalous mass enhancement in SVO QW states, and the contribution is remarkably enhanced as $E_n$ decrease.

The trend of the observed enhancement of electron correlation as the subband bottom energy approaches $E_\mathrm{F}$ seems to be opposite to what would be expected from the simple Fermi-liquid behavior of bulk SVO. 
In general, the strength of the electron-electron interaction is proportional to the square of the density of states $D(\omega)$ at $E_\mathrm{F}$, $\left[ D \left( E_\mathrm{F} \right) \right]^2$, and similarly the inverse lifetime of the impurity scattering is proportional to $D \left( E_\mathrm{F} \right)$. 
Therefore, based on the behavior of bulk SVO \cite{PRB_98_Inoue}, a weaker correlation effect is expected in the subbands located at lower binding energies owing to the reduction of band filling \cite{PRB_70_Brinkman}. 
The seeming contradictions can be reconciled by considering the orbital-selective quantization of the V $3d$ $t_{2g}$ states involving the 2D nature of the subbands in each respective plane. 
In the present case, the quasi-1D electronic states are realized in the $d_{yz}/d_{zx}$-derived subbands by the further dimensionality reduction (from 2D to 1D) due to the quantum confinement. 
As a result, the characteristic orbital-dependent quantized $t_{2g}$ band states cause the FSs to consist essentially of some pairs of roughly parallel lines along the $k_{x}/k_{y}$ directions \cite{Science_11_Yoshimatsu} originating from the quasi-1D $d_{yz}/d_{zx}$ states and a bulk-like circular FS from the $d_{xy}$ states (maintaining their 2D nature), as can be seen in Fig.~\ref{hvDep}(d). 
If electron correlation in such a situation is dominated by electron-electron scattering within each subband, one has to consider the density of states of each 1D subband, and not the density of states of the entire V $3d$ $t_{2g}$ band as in bulk SVO.

In 1D systems, the density of states is given by $D(E) \propto 1/\sqrt{E}$, where $E$ is the energy measured from the band edge (the subband minimum in the present case), being different from $D(E) = \mathrm{const.}$ in 2D and $D(E) \propto \sqrt{E}$ in 3D. 
Thus, the scattering rate $\beta_n$ within the quasi-1D $d_{yz}/d_{zx}$ subband will be proportional to the square of the density of states at $E_\mathrm{F}$ and hence will be enhanced by a factor of $1/E_n$. 
In order to test the validity of the enhancement of electron correlation derived from the quasi-1D subband, a function proportional to $1/E_n$ is fitted to the plot of the coefficient $\beta_n$ against $E_n$, as shown in Fig.~\ref{ImS_beta}(d). 
The curve fit adequately reproduces the experimental results. 
The validity is further confirmed by the $\Gamma_n^\mathrm{imp}$ data, which should be proportional to the density of states at $E_\mathrm{F}$. 
Indeed, $\Gamma_n^\mathrm{imp}$ is well scaled with $1/\sqrt{E_n}$, as shown in Fig.~\ref{ImS_beta}(e). 
These good agreements strongly suggest that the anomalous subband-dependent mass renormalization is affected by the 1D character of the confined V $3d$ subband states as a result of selective quantization of orbitals involving the 2D nature \cite{RKW}. 

The present conclusion is valid if the electron-electron scattering is the strongest within each subband and the inter-subband electron-electron scattering is weaker. 
This condition will be satisfied because the Coulomb interaction should be stronger between electrons in the same QW band, i.e., with the same quantum number $n$ owing to the electrons having the same wave functions in the $z$ direction.
In fact, it is evident from Fig.~\ref{ImS_beta} that the $\mathrm{Im}\Sigma_n (\omega)$ curve of the $n=1$ state remains almost unchanged with the increase in thickness from 6 to 8 ML, even when the new subband of the $n = 3$ state emerges at 7 ML and it energetically approaches other states at 8 ML. 
The absence of any detectable changes in the $n=1$ states despite the emergence of a new subband and/or the approach to other subbands indicates that the inter-subband interaction is too weak to be observed within the given experimental accuracy.

In conclusion, we have addressed the origin of the anomalous subband-dependent mass enhancement observed in SVO QW structures using {\it in situ} ARPES measurements. 
The detailed line-shape analysis reveals that the obtained $\mathrm{Im} \Sigma(\omega)$ significantly increases as the subband bottom energy approaches $E_\mathrm{F}$, consistent with the subband dependence of the renormalization factor $Z_n$. 
Assuming the Femi-liquid behavior of the SVO QW states, these results strongly suggest that the electron correlation is dominated by intra-subband electron-electron scattering and significantly affects the anomalous subband-dependent mass renormalization observed in SVO metallic QW structures. 
Taking into account the characteristic orbital-selective quantization, we have concluded that the enhancement of electron-electron correlation originates from the quasi-1D electronic structure of the quantized states, where the density of states diverges at the band bottom. 
The present study demonstrates that the QW structure of strongly correlated oxide will provide a platform for both investigating the behavior of correlated electrons under varying interactions among their spin, charge, and orbital degrees of freedom and for manipulating novel quantum phenomena in reduced dimensions.

The authors are very grateful to S. Okamoto and M. Minohara for useful discussions. 
This work was supported by a Grant-in-Aid for Scientific Research (B25287095 and S22224005), a Grant-in-Aid for Young Scientists (26870843), and a Research Activity Start-up grant (25887021) from the Japan Society for the Promotion of Science (JSPS) and the MEXT Elements Strategy Initiative to Form Core Research Center. 
This work at KEK-PF was performed under the approval of the Program Advisory Committee (Proposals 2012G536, 2012G668, and 2013S2-002) at the Institute of Materials Structure Science at KEK.


\begin{thebibliography}{10}
\bibitem{M.E.}
Changes in the ARPES intensity at the equivalent symmetric points may 
  originate from matrix-element effects and/or asymmetries in the measurement
  geometry.
\end{thebibliography}


\end{document}